\documentclass[final,5p,times,twocolumn]{elsarticle}

\usepackage{amssymb}
\usepackage[utf8]{inputenc} 
\usepackage[T1]{fontenc}    
\usepackage{hyperref}       
\usepackage{url}            
\usepackage{booktabs}       
\usepackage{amsmath}
\usepackage{amssymb}
\usepackage{siunitx}
\usepackage{nicefrac}       
\usepackage{microtype}      
\usepackage{lipsum}
\usepackage[caption=false]{subfig}
\usepackage{dcolumn}
\usepackage{bm}
\usepackage{microtype}
\usepackage[capitalize]{cleveref}
\usepackage{listings}
\usepackage[mathlines]{lineno}

\newcommand{\rff}{r_{\text{eff}}}
\newcommand{\zff}{z_{\text{eff}}}
\newcommand{\zen}{\Theta_{\text{zen}}}
\newcommand{\azi}{\Phi_{\text{azi}}}

\journal{Nuclear Instruments \& Methods in Physics Research, Section A}

\begin{document}

\begin{frontmatter}



\title{Detecting neutrinos in IceCube with Cherenkov light in the South Pole ice}


\author{Tianlu Yuan \textit{for the IceCube collaboration}}

\affiliation{organization={Dept.~of Physics and Wisconsin IceCube Particle Astrophysics Center },
            addressline={University of Wisconsin -- Madison}, 
            city={Madison},
            postcode={53706}, 
            state={WI},
            country={USA}}

\begin{abstract}
The IceCube Neutrino Observatory detects GeV-to-PeV+ neutrinos via the Cherenkov light produced by secondary charged particles from neutrino interactions with the South Pole ice. The detector consists of over 5000 spherical Digital Optical Modules (DOM), each deployed with a single downward-facing photomultiplier tube (PMT) and arrayed across 86 strings over a cubic-kilometer. IceCube has measured the astrophysical neutrino flux, searched for their origins, and constrained neutrino oscillation parameters and cross sections. These were made possible by an in-depth characterization of the glacial ice, which has been refined over time, and novel approaches in reconstructions that utilize fast approximations of Cherenkov yield expectations.

After over a decade of nearly continuous IceCube operation, the next generation of neutrino telescopes at the South Pole are taking shape. The IceCube Upgrade will add seven additional strings in a dense infill configuration. Multi-PMT OMs will be attached to each string, along with improved calibration devices and new sensor prototypes. Its denser OM and string spacing will extend sensitivity to lower neutrino energies and further constrain neutrino oscillation parameters. The calibration goals of the Upgrade will help guide the design and construction of IceCube Gen2, which will increase the effective volume by nearly an order of magnitude.
\end{abstract}




\end{frontmatter}


\section*{Introduction} \label{sec:intro}
\begin{figure}[hbt]
\centering
\includegraphics[width=0.9\columnwidth]{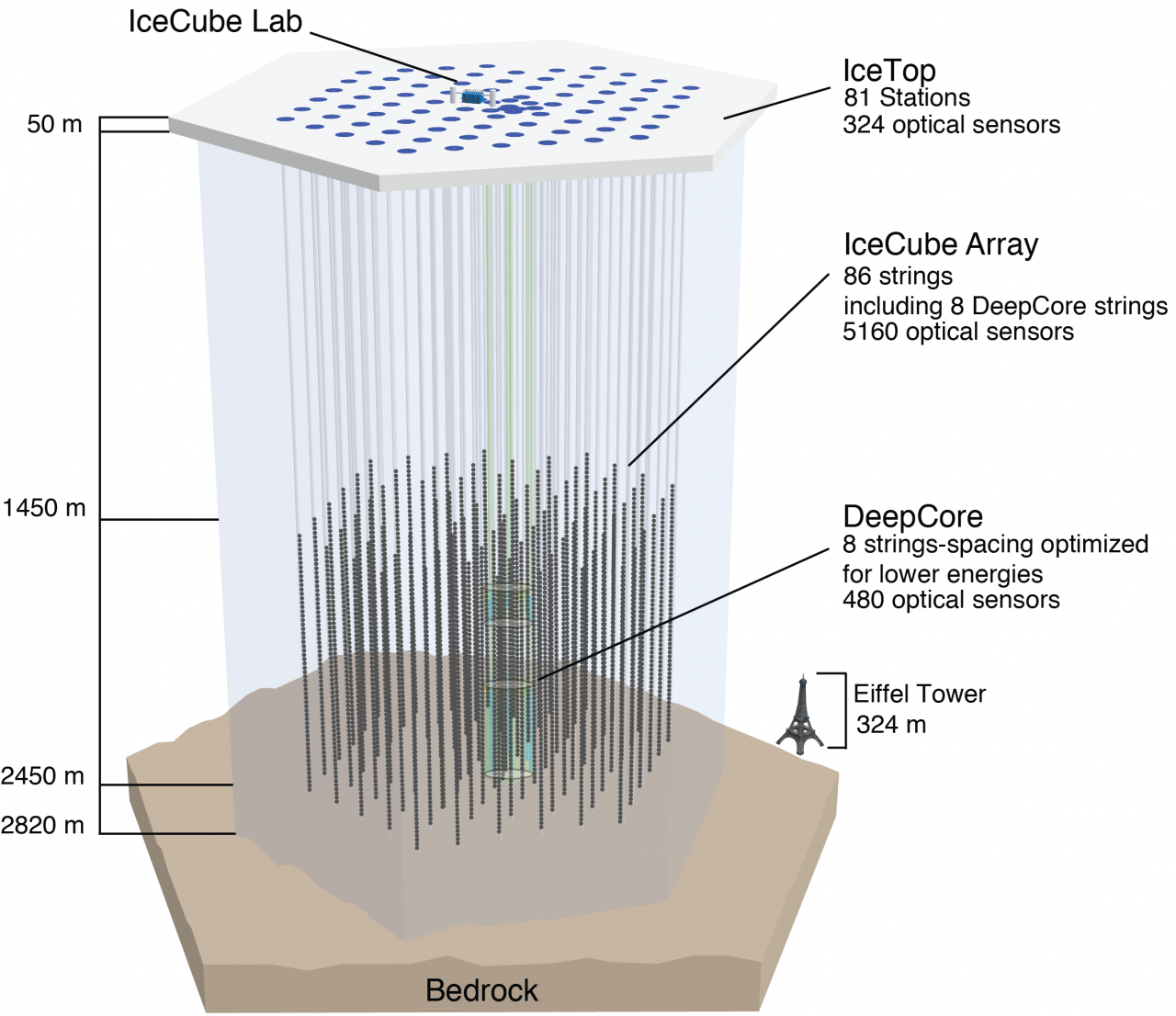}
\caption{Schematic of the IceCube Neutrino Observatory~\cite{Aartsen:2016nxy}. DeepCore is highlighted in the center of the detector.} 
\label{fig:detector}
\end{figure}

The IceCube Neutrino Observatory detects neutrinos interacting with nucleons and electrons in the South Pole ice via Cherenkov radiation produced by charged secondaries. It is instrumented with 5160 Digital Optical Modules (DOM), each with a single downward-facing photomultiplier tube (PMT), arrayed across a cubic kilometer. The DOMs are attached to 86 strings --- cables installed in the ice that provide mechanical and electrical support. DOMs are spaced \SI{17}{\m} apart on standard IceCube strings and \SI{7}{\m} apart on DeepCore strings, a denser infill region of the detector. Standard IceCube strings are spaced approximately \SI{125}{\m} apart. \Cref{fig:detector} illustrates the scale and hexagonal configuration of the in-ice detector.

IceCube produces a rich and diverse physics program probing the largest and smallest scales. Highlights include the first discovery and later confirmation of a diffuse flux of astrophysical neutrinos~\cite{IceCube:2013low,IceCube:2020wum}, the first identification of an astrophysical neutrino source TXS 0506+056~\cite{IceCube:2018dnn} arising from its realtime program~\cite{IceCube:2016cqr}, and the detection of the first astrophysical neutrino interaction on the Glashow resonance~\cite{IceCube:2021rpz}. Most recently, the nearby Seyfert galaxy NGC 1068 has also been identified as a possible steady source of astrophysical neutrinos~\cite{IceCube:2022der}. These and future results rely on, and will continue to benefit from, refined calibration of the detector and improved event reconstruction.

\section*{Ice anisotropies and calibration} \label{sec:ice}
Accurate characterization of physics quantities across a sparse array of PMTs requires accurate calibration of ice and instrument. This has been accomplished with calibration LEDs attached onto a dedicated flasher board of each IceCube DOM, and using large samples of downgoing minimum ionizing muons to set the global energy scale. The propagation of light in the glacial ice sheet is complex at Cherenkov wavelengths, with dependencies on the depth that reflect Earth's climate across geological timescales~\cite{IceCube:2013llx}. Absorption and scattering coefficients that describe the mean free path of a photon are fitted using calibration LED data across all instrumented depths. In general, the absorption (scattering) length in ice is longer (shorter) than in water Cherenkov detectors. IceCube also discovered a directional dependence in the light propagation, or ice anisotropy, with maximal effect along and perpendicular to the ice flow axis~\cite{Chirkin:2013}. A microscopic explanation of the ice anisotropy due to birefringence of polycrystals was proposed in~\cite{Chirkin:2019vyq} and yields improved agreement with calibration data over the previous phenomenological model~\cite{tc-2022-174}. Though the instrumentation is too sparse for imaging Cherenkov rings, refinements of ice properties translate into more accurate simulations and reconstructions.

\begin{figure}[hbt]
\centering
\includegraphics[width=0.9\columnwidth]{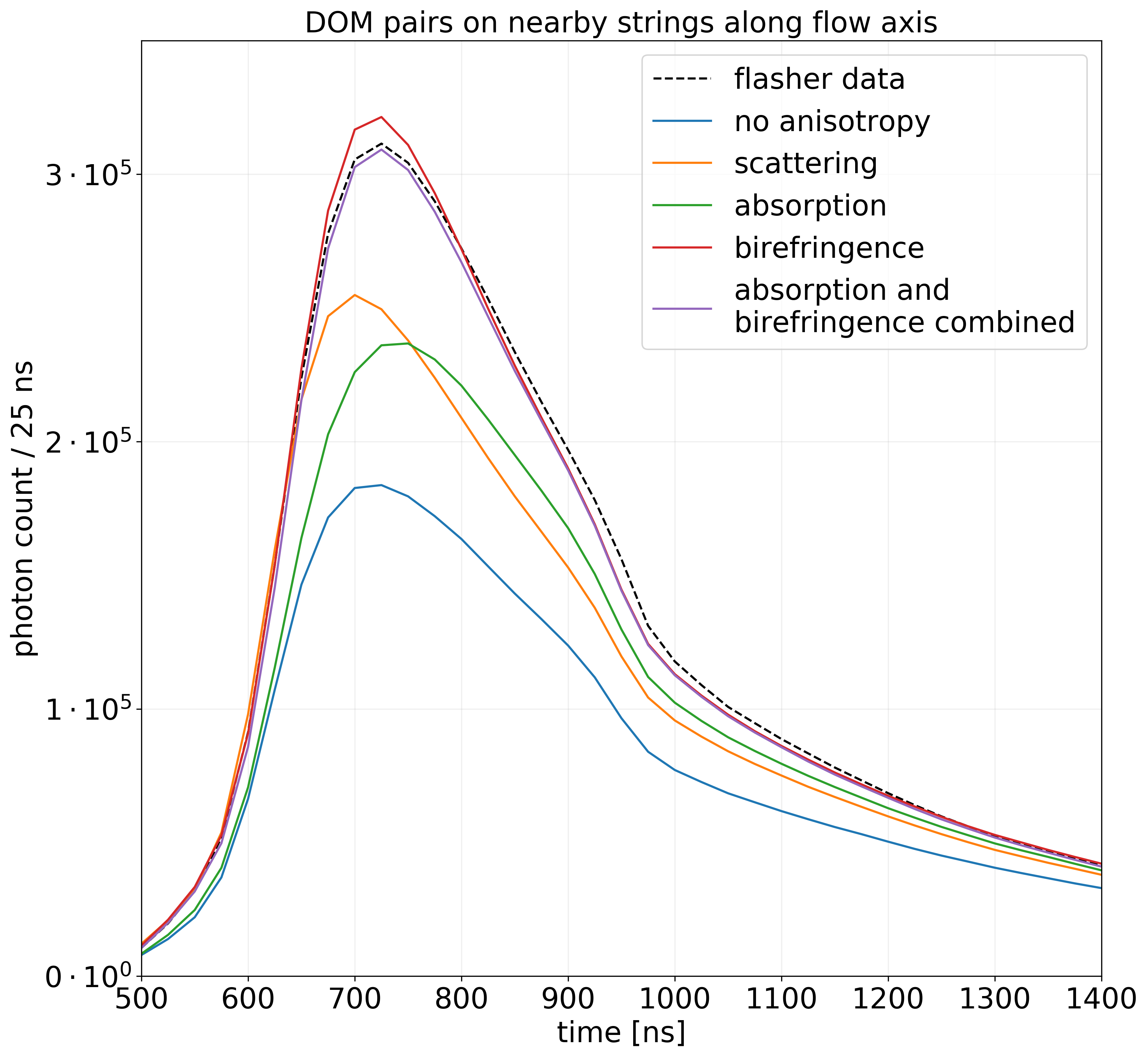}
\caption{Figure taken from~\cite{tc-2022-174} which show photon arrival time distributions from data (black) and a series of different ice models (colored lines) in \SI{25}{\nano \s} bins. A model that combines the effect of birefringence and anisotropic-absorption (purple) yields the best agreement with data to date.}
\label{fig:flow}
\end{figure}
To exploit the massively parallel computation nature of photon propagation, IceCube relies on GPUs for a significant part of its simulation chain~\cite{Chirkin:2013tma,Chirkin:2019rcj}. Ice properties can be measured and refined by simulating calibration LED devices and comparing to the observed calibration data~\cite{IceCube:2013llx}. The impact of recent improvements to the ice model can be seen in \cref{fig:flow}, which shows a comparison of photon arrival time distributions for nearby emitter-receiver pairs roughly oriented along the ice flow. The expected distributions from simulation are also shown for four ice models implementing different anisotropy modeling, as well as one without anisotropy (blue). The best description with data (black) to-date is obtained with a combination of birefringence and anisotropic-absorption (purple)~\cite{tc-2022-174}. It is worth noting the striking improvement when including the microscopic birefringence effect (red, purple) over previous models, which implement an effective parametrization of the anistropy in terms of modifications to the scattering or absorption lengths, or no anisotropy at all. A distinguishing feature of birefringence models (red, purple) is that the probability density in time does not shift as much as the previous effective parameterizations (green, orange) relative to the no-anisotropy model (blue).

\begin{figure*}[hbt]
\centering
\includegraphics[width=0.32\textwidth]{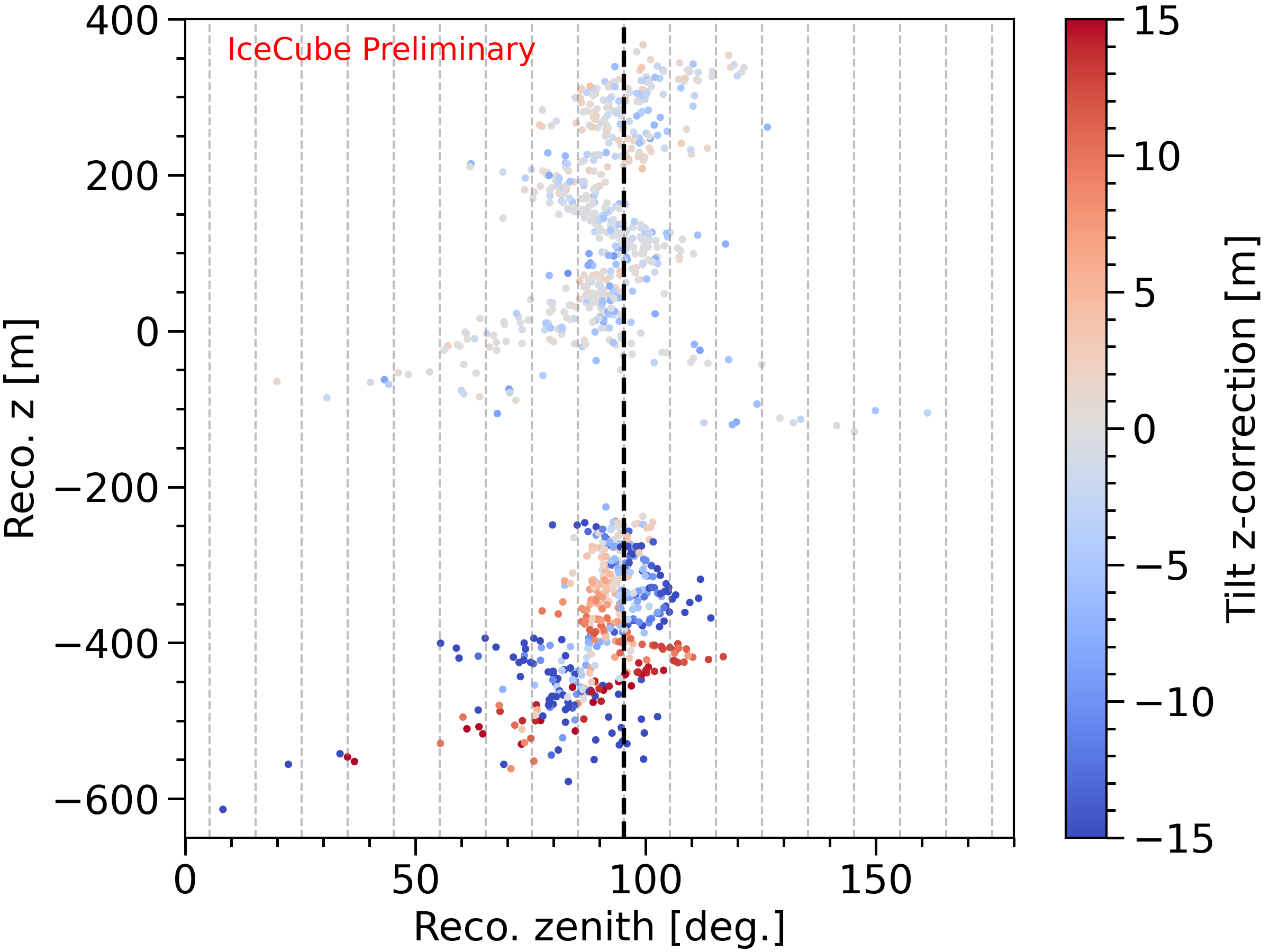}
\includegraphics[width=0.32\textwidth]{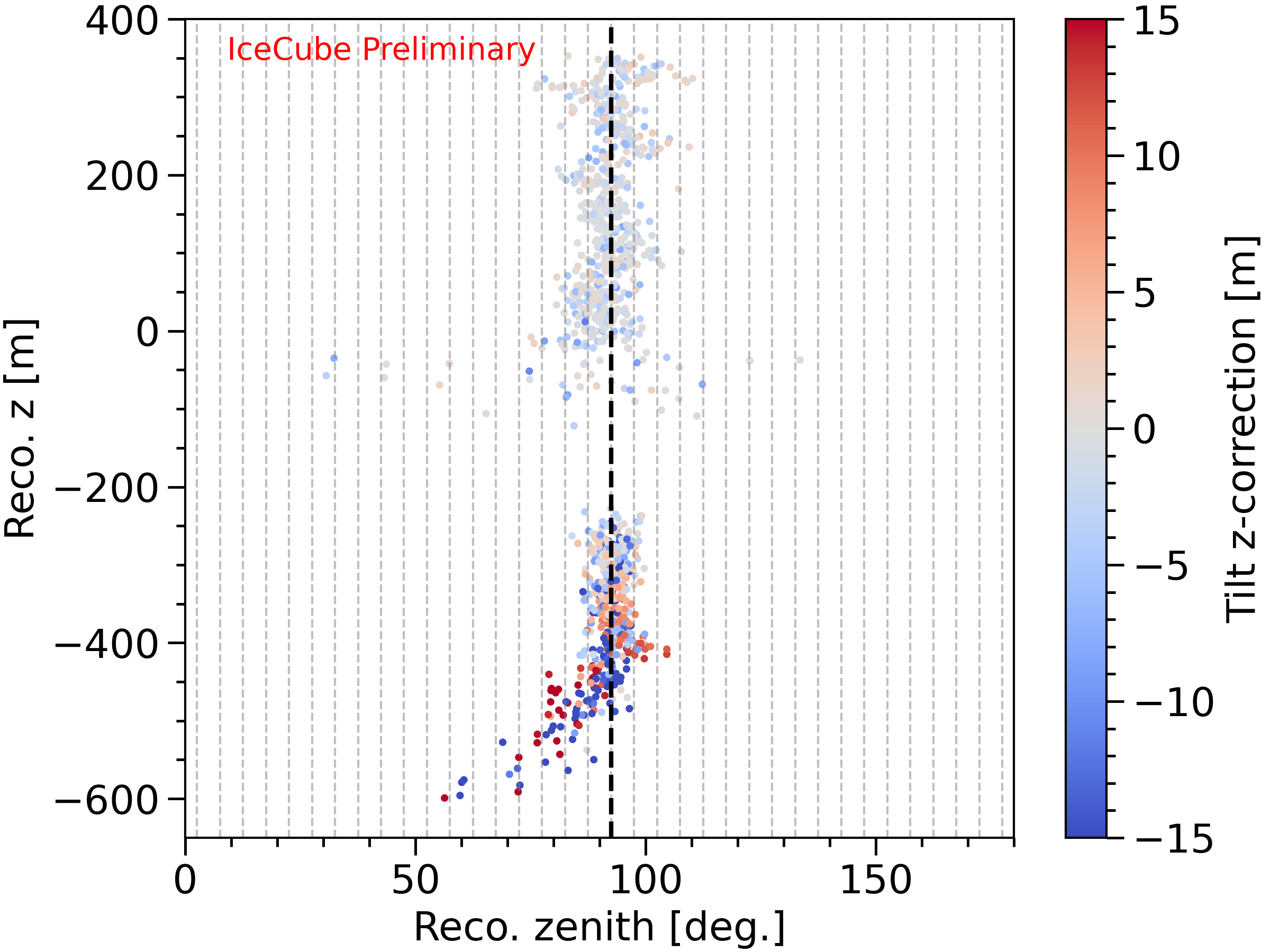}
\includegraphics[width=0.32\textwidth]{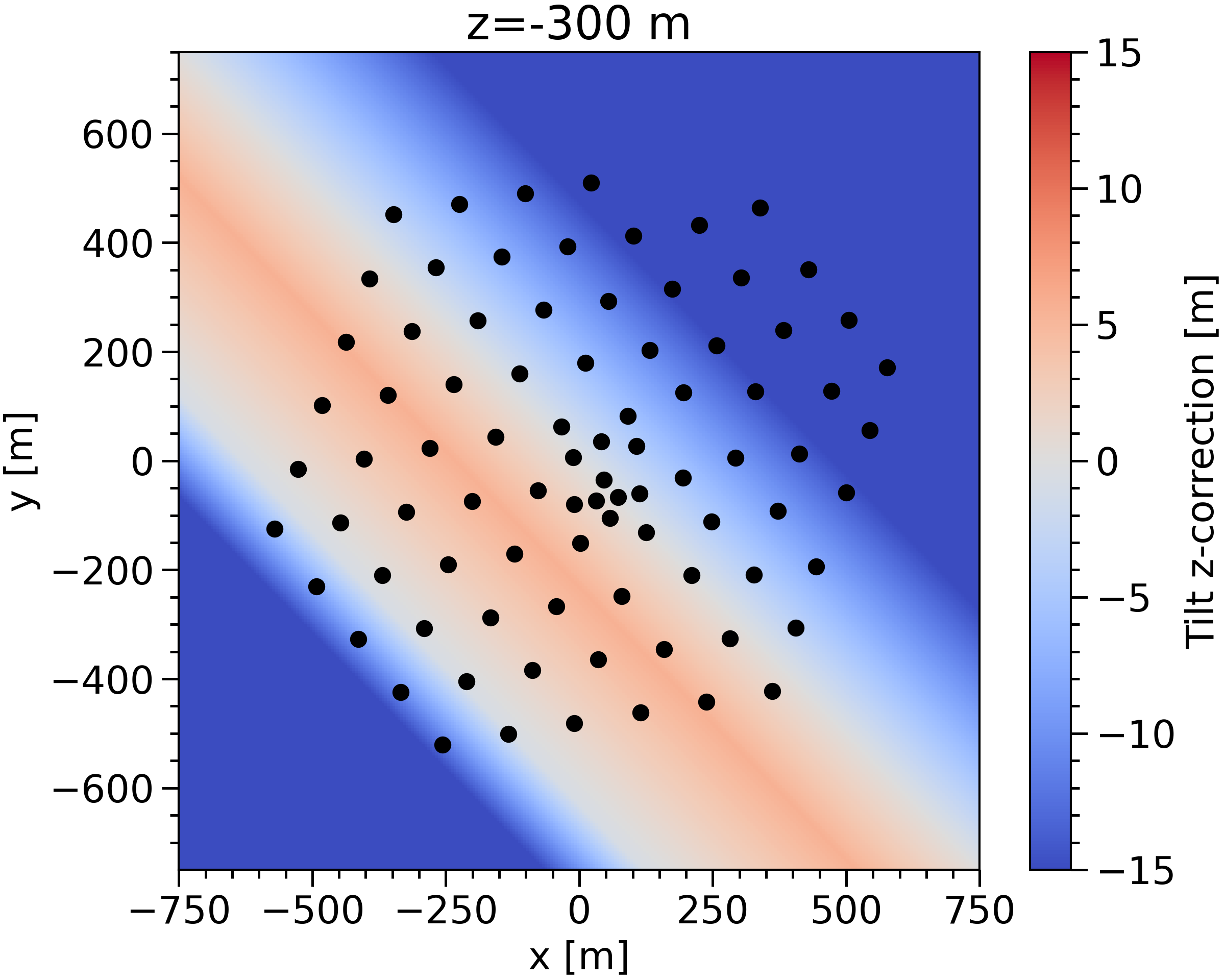}
\caption{Scatter plots of reconstructions for a set of simulated cascades with true zenith near the black dashed line. The cascades were simulated inside the detector using a model with ice tilt, and a clear bias in the reconstructed zenith is visible when reconstruction assumes perfectly horizontal ice layers (left panel). This bias is largely mitigated when using an updated model for reconstruction that includes approximations to ice anisotropies (center panel). The grey dashed lines indicate B-spline knot positions in $\zen$. The gap near $z=-190$ is due to a layer of heightened scattering and absorption across the detector, and the tail at the lowest depths are events that reconstructed below the active region of the detector. The right panel illustrates the simulated tilt direction at a depth of $z=\SI{-300}{\m}$ in detector coordinates and black dots indicate string $(x,y)$ positions. For all three panels, the color indicates the tilt z-correction, $\zff-z$, in meters. See text for more details.}
\label{fig:tilt}
\end{figure*}
Further, it is well known that ice properties change as a function of vertical depth over the kilometer scale that IceCube spans. Layers of relatively homogeneous ice isochrons each have their unique set of scattering and absorption coefficients. Using data from a dust logger deployed in seven IceCube boreholes, the layers were discovered to exhibit undulations~\cite{IceCube:2013jrb}. These layer undulations, or ice tilt, introduce an effective depth for each layer at each point across the detector, defined relative to a reference location where the stratigraphy is correct in terms of the true depth. This can be seen in the right panel of \cref{fig:tilt}, which illustrates the effective tilt correction in $z$ across a horizontal slice at a true depth of $z=\SI{-300}{\m}$ (detector coordinates), with $(x,y)$ locations of the 86 IceCube strings shown in black dots. The left and middle panels show  scatter plots of the reconstructed zenith direction (x-axis) vs depth (y-axis) for a set of simulated cascade (particle shower) events with true zenith within one degree of the black dashed line. The simulation includes our best knowledge the ice tilt at the time, and the color indicates the depth-correction needed to properly account for this effect at each cascade position. A bias correlated with the tilt depth-correction is visible in the left panel, which assumes flat ice layers in reconstruction, and disappears for a reconstruction that incorporates tilt (middle panel). Additional improvements independent of tilt are also included in the middle panel and will be the focus of the next section.

\section*{Approximative models for reconstruction} \label{sec:model}
Processing speed on GPUs is often sufficient for simulation, but reconstruction can require orders of magnitude more trials for each event. Fast approximations of Cherenkov yields have been developed to address this, initially based on look-up tables, and now using a tensor product of B-splines~\cite{Aartsen:2013vja, Whitehorn:2013nh} and neural networks (NN)~\cite{IceCube:2021umt}, derived from simulations of cascades or muon-induced tracks. Accurate ice modeling is especially important for cascades, which approximate not only electron- or hadron-induced particle showers but also stochastic energy-loss processes in high-energy muons above \SI{1}{\tera \eV}. New ice model features should be included in the approximators for best results. 

Anisotropies in the ice due to effects like the birefringence and layer undulations bring additional complexity by breaking symmetry and thus introducing additional dimensionality. A full parametrization of the photon yield for cascades would span 9 dimensions: $(x,y,z)$ to define the cascade position, $(\zen, \azi)$ to define its direction, $(r, \theta, \phi)$ to define the DOM position relative to the cascade, and $t$ the photon-arrival time at the DOM. This increased complexity makes a tensor product of B-splines difficult, if not intractable, to evaluate over all relevant regions of the parameter space. In addition, the simulation needed to construct approximators becomes prohibitively expensive to compute. While NN are not bound by dimensionality constraints, they can be subject to unknown inaccuracies in interpolation and extrapolation while also requiring a large simulated dataset.

\begin{figure}[hbt]
\centering
\includegraphics[width=0.9\columnwidth]{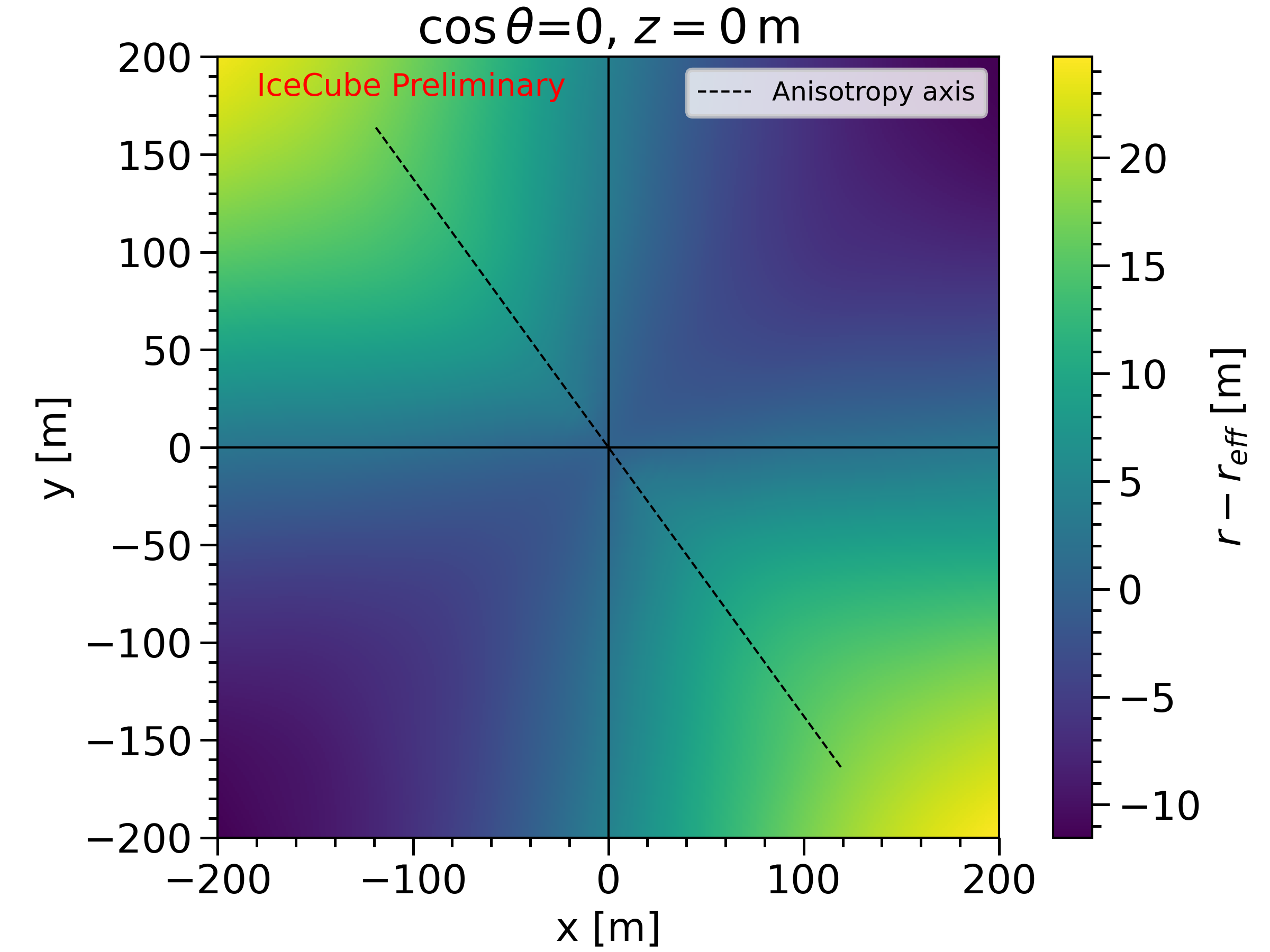}
\caption{A visualization of the ice anisotropy due to birefringence across the center of the detector. The color indicates $r-\rff$, with larger (smaller) values corresponding to regions where the expected photon yield is higher (lower). The dashed line corresponds to the axis of ice flow. The effect shown in \cref{fig:flow} is visible, where a higher photon yield relative to the no-anisotropy model is expected along the flow axis.}
\label{fig:reff}
\end{figure}
An alternative approach, originally proposed in~\cite{Usner:2018cel}, is to treat the ice anisotropy and tilt separately from a bulk ice that assumes flat, horizontal layers without anisotropy. Modeling each of these independently restores some symmetry and reduces the dimensionality for cascades to $(r, \theta, \phi, t, \zen, z)$, while also isolating the effect due to ice. The anisotropy and tilt can be modeled with a function $\rff (r, \theta, \phi, x, y, z)$, where $(r, \theta, \phi)$ define the receiver location relative to an \emph{isotropic} light source\footnote{The isotropic source simplifies the problem as there is no need to include directional information of the particle shower.} in spherical coordinates and $(x,y,z)$ defines the source position. Initially, however, this was simplified by assuming a global approximation of tilt $\rff'(r, \theta, \phi, z) = \rff(r, \theta, \phi, 0,0,z)$ such that, as a rough estimate, only a source $z$-dependence remained~\cite{Usner:2018cel}. Two simulation sets are produced, one assuming a flat ice model without anisotropy, and one assuming the complete description of the ice. An isotropic source can then be used to model effects due solely to the ice; for a given $r$ any differences between the simplified ice model and the full description can be attributed to effects induced by the ice~\cite{Usner:2018cel}. As the expected photon yield falls off monotonically with $r$ it is possible to construct a mapping such that the time-integrated photon yield $N_{\text{simple}}(\rff'(r, \theta, \phi, z)) = N_{\text{complete}}(r, \theta, \phi, z)$ for an isotropic source~\cite{Usner:2018cel}. A tensor-product B-spline is then fit to simulated data to construct a smooth approximator to $\rff$.

We here suggest that it is more correct to factorize the tilt out from $\rff$ so that only the ice anisotropy due to birefringence is modeled by $\rff(r, \theta, \phi, z)$. Birefringence applies globally across the detector, independent of $(x,y)$. The $z$-dependence cannot be removed due to the layer-by-layer differences in the ice. Again, two simulation sets are produced, one assuming a flat ice model without anisotropy and another with only anisotropy, such that the time-integrated yields $N_{\text{simple}}(\rff(r, \theta, \phi, z)) = N_{\text{anis}}(r, \theta, \phi, z)$. The obtained $r-\rff$ is visualized for a slice across $\cos \theta=0$ and $z=\SI{0}{\m}$ (center of IceCube) in \cref{fig:reff}. Brighter (darker) regions indicate where a higher (lower) photon yield is expected in the anisotropic ice model relative to the simplified model. 

The tilt correction is separately approximated by directly reading in tabulated layer effective depths and linearly interpolating to obtain a tilt-corrected depth $\zff(x,y,z)$. This approach is taken in the full photon simulation as well, albeit at the individual photon scattering position rather than a single source position. The photon yields for cascades can then be obtained by substituting $\rff$ and $\zff$ for $r$ and $z$, respectively. With the birefringence treatment of ice anisotropy, it was found that the probability density in time remains relatively unchanged, so $\rff$ is only applied to rescale the amplitude of the photon yields for approximating birefringence.

Finally, fitting multidimensional B-splines using a tensor product construction can lead to interpolation artifacts along diagonals, and extrapolation failures near boundaries if boundary conditions are not enforced. The former is simply due to defining basis functions on cartesian grids~\cite{Khalil}; akin to failures in bilinear interpolation when strong features lie across two points along a diagonal. In particular this effect led to biases in the reconstructed zenith direction, visible in the left panel of \cref{fig:tilt} as a clustering of points in between two knot positions (dashed lines) in $\zen$. An unbiased reconstruction should be centered on the true zenith (dashed black line), but in the left panel we see that reconstructed directions tend to distribute away from the truth. This was observed with the previously generated splines even when an identical ice model is used in the simulation, and hinted at biases in the construction of the approximative model itself. The effect arises from interpolation artifacts along the $(\theta, \zen)$ diagonal. Simulating a finer set of support points along $\zen$ and updating the knot placements removes this bias as shown in the right panel of \cref{fig:tilt}.

Extrapolation failures can also occur when the fitted spline extends beyond the last support point. These failures led to sharp discontinuities in the photon yields across boundaries, and occurred in the spherical coordinates system defined by $\theta$ and $\phi$, near the poles where $\cos \theta = \pm 1$ and near the boundary of $\phi=0$. The former required the construction of an additional data point at $\cos \theta = \pm 1$ based on linear interpolation, and the later was resolved by reflection across $\phi = 0$. By including these additional support points, boundary discontinuities were largely reduced.

\begin{figure}[hbt]
\centering
\includegraphics[width=0.9\columnwidth]{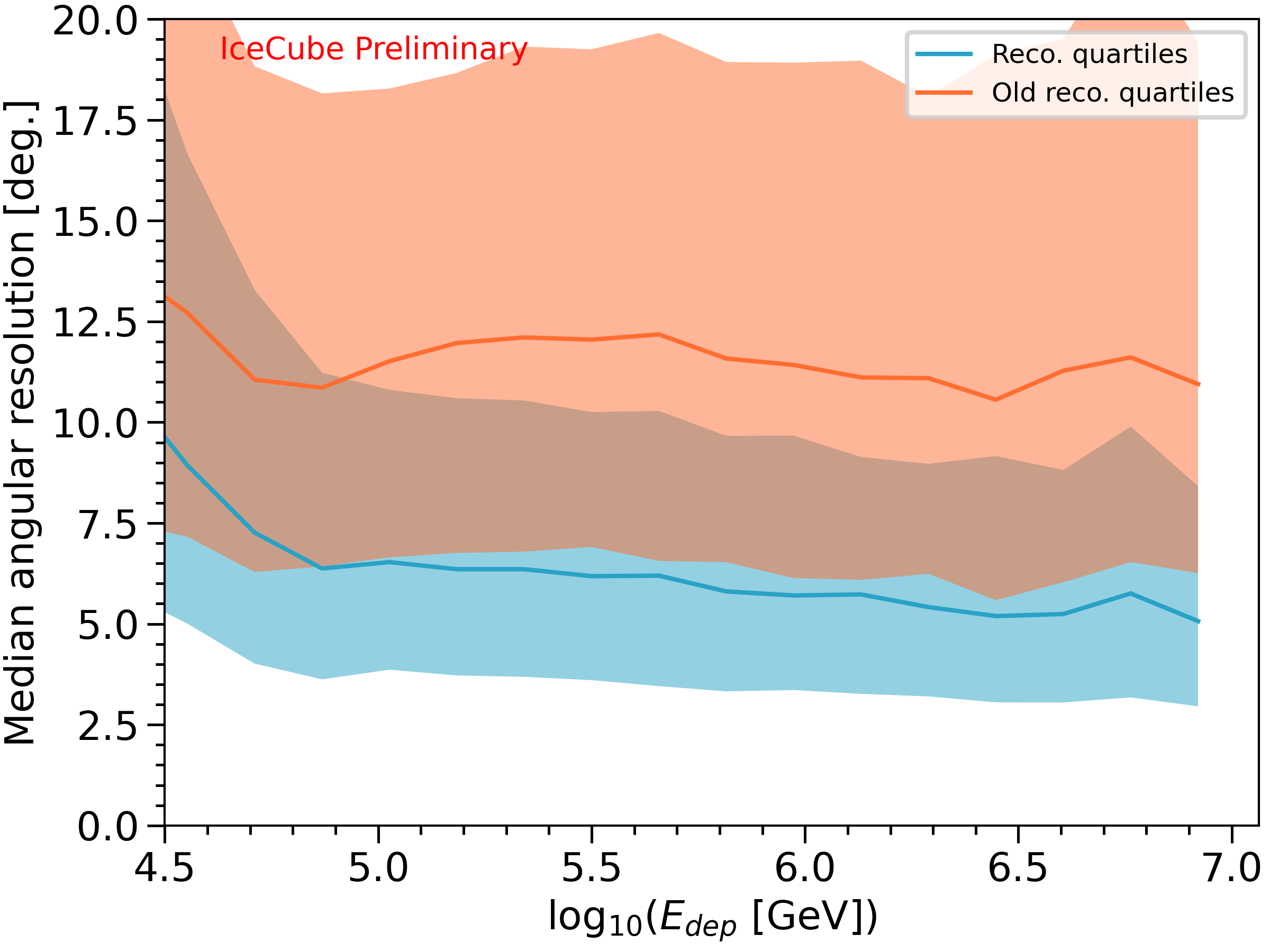}
\caption{Quartiles of the distribution between reconstructed and true directions for a sample of simulated, contained cascades as a function of deposited energy. The angular resolution improves by a factor of two using the newer model that includes new ice features as approximated by $\rff$ and $\zff$, as well as updates to the B-spline fits.}
\label{fig:res}
\end{figure}
By accounting for ice anisotropy with $\rff$ and tilt with $\zff$, and reducing interpolation and extrapolation failures, the approximative model used in cascade reconstruction can be shown to give a much better description of the ice. This is highlighted in \cref{fig:res} which shows the angular resolution -- or angular distance between the reconstructed direction and the true direction -- for a sample of simulated, contained cascades as a function of energy deposited in the detector. Going from an older model (orange) to the updated model (blue), the angular resolution quartiles improve by a factor of two.

\section*{Outlook} \label{sec:conclusion}
As discussed in this proceeding and elsewhere, much recent progress has been made in the context of improved detector calibration and its approximative models used in reconstruction. However, the angular resolutions shown here are still far from the statistical limit under ideal conditions. The limitations lie in the simplifications and assumptions that were made, and additional smearing introduced by the PMT and electronics. Even if these challenges are not entirely resolved, the technical knowledge gained in attempts to better model them will serve to inform preparations for the next generation of neutrino telescopes at the South Pole. 

The IceCube Upgrade adds seven new strings in the DeepCore infill region~\cite{Ishihara:2019aao}. New, multi-PMT OMs will be attached with denser spacing to each string and increases the effective coverage of each OM. With its lower energy reach, the Upgrade will be able to precisely measure the rate of tau neutrino appearance and probe if the Pontecorve-Maki-Nakagawa-Sakata matrix that governs neutrino oscillations is unitary. In addition, the Upgrade will deploy new calibration devices to constrain remaining ice uncertainties.

To improve upon measurements of the astrophysical neutrino flux and pinpoint their sources, the IceCube Gen2 detector will add 120 strings to the ice surrounding the existing IceCube detector and increase the target volume by nearly an order of magnitude~\cite{Aartsen:2020fgd}. IceCube Gen2 will include multi-PMT OMs that are currently under design, and extend sensitivities to neutrino energies above 10 PeV.



\bibliographystyle{elsarticle-num} 
\bibliography{main}
\end{document}